\documentclass[pra,twocolumn,showpacs]{revtex4}  % twocolumn preprint
\usepackage{graphicx}
\usepackage{amssymb}

\begin{document}

\title{Observation of triply coincident nonlinearities in periodically poled $\mathbf{KTiOPO_4}$} 
\author{Raphael C. Pooser}
\author{Olivier Pfister}
\email[Corresponding author: ]{opfister@virginia.edu}
\affiliation{Department of Physics, University of Virginia, 382 McCormick Road, Charlottesville, VA 22904-4714, USA}
\begin{abstract}
We report the simultaneous quasi-phase-matching of all three possible nonlinearities for propagation along the $X$ axis of periodocally poled (PP) $\rm KTiOPO_4$ (KTP) for second-harmonic generation of 745 nm pulsed light from 1490nm subpicosecond pulses in a PPKTP crystal with a 45.65 $\mu$m poling period. This confirms the recent Sellmeier fits of KTP by K. Kato and E. Takaoka \protect{[\ao {\bf 41}, 5040 (2002)]}. Such coincident nonlinearities are of importance for realizing compact sources of multipartite continuous-variable entanglement \protect{[Pfister {\em et al.},\pra {\bf 70}, 020302 (2004)]} and we propose a new simpler method for entangling four fields, based on this triple coincidence.
\end{abstract}
\maketitle

One of the interests of continuous-variable (CV) quantum information (QI) is its relative ease of implementation by use of the well established experimental techniques of  quantum optics.\cite{book,rmp} Recently, a quantum teleportation network was realized out of CV multipartite entangled states,\cite{furusawa} as originally proposed by van Loock and Braunstein.\cite{vLB} In such an achievement, the quantum entangled channel uses 3 different optical parametric oscillators (OPO) below threshold, whose output squeezed beams are made to stably interfere with one another. In the limit of strong squeezing, such states are CV analogs of GHZ states.\cite{vLB} Recently, we have shown theoretically that a more compact setup, using but a single OPO, can generate the exact same states if the OPO's nonlinear medium provides simultaneous phase-matching of two different nonlinearities (concurrence).\cite{us}
Such coincidences are extremely difficult to realize with birefringent phase-matching (BPM) but become straightforward with quasi-phase-matching (QPM) in, say, periodically poled (PP) ferroelectrics.\cite{ol97} In this Letter, we present the experimental realization of the nonlinear coincidence suitable for the entanglement generation proposed in Ref.~\cite{us}, in a PP-$\rm KTiOPO_4$ (KTP) crystal.

\begin{figure}[htb]
\begin{center}
\includegraphics[width=3.25in]{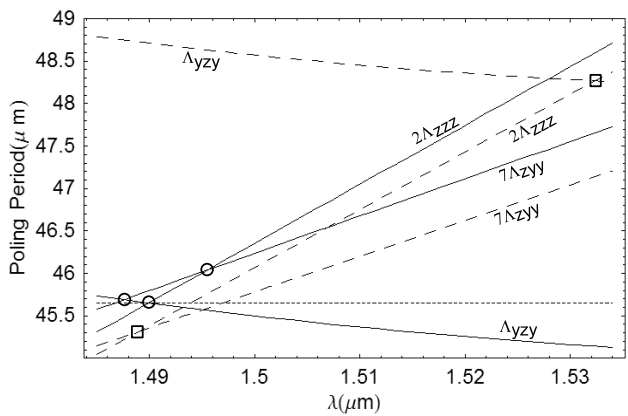}
\end{center}
\caption{Theoretical QPM-SHG solutions for $\Lambda_{YZY}$, $2\Lambda_{ZZZ}$, and $7\Lambda_{ZYY}$, which are the only interactions corresponding to nonzero nonlinear tensor  elements in KTP. All other harmonics are out of range. The $n^{th}$ subharmonic has an effective nonlinear coefficient\cite{fejer} $2d/\pi n$, hence we sought the lowest orders.  The dashed curves 
represent the poling periods calculated using the Sellmeier equations from Emanueli {\em et al.} \cite{emanueli} while the solid curves represent poling periods calculated using Selmier equations from Kato  {\em et al.} \cite{kato}.  The horizontal dotted line represents the chosen poling period of the PPKTP crystal at 45.65$\mu$m.}
\label{qpm}
\end{figure}

Since the coming of age of reliable periodic poling techniques in ferroelectric nonlinear crystals such as $\rm LiNbO_3$,\cite{myers} the QPM technique\cite{qpm} has become accessible. It relies on the spatial modulation of the nonlinear coefficient induced by periodic poling, rather than on birefringence, to compensate for the phase mismatch of any given nonlinear interaction. As a consequence, the same nonlinear material can be used to phase-match a wide range of nonlinear optical mixings by simply changing the poling period, with noncritical QPM conditions (no walkoff) always ensured. Moreover, the larger, diagonal (polarization degenerate) nonlinear tensor elements can be used, contrary to the case of BPM, which greatly increases the nonlinear conversion efficiency. Last but not least, sophisticated nonlinear systems can be made by fabricating multi-period crystals or Fourier engineering the spatial modulation.\cite{fejer} Initially achieved in $\rm LiNbO_3$, periodic poling has also been successfully obtained in RTA \cite{pprta} and KTP \cite{ppktp}, which is of great interest for CV quantum optics as these crystals present reduced linear and nonlinear absorption and photorefractive damage. However, even though PPKTP is commercially available, the knowledge of its Sellmeier coefficients is not yet quite at the level required for designing QPM concurrences with full confidence. For this reason, we decided to keep the design simple, with but a single poling period, thereby allowing us to also characterize the material further. As proposed in Ref.~\cite{us}, we sought to implement the concurrence of ZZZ and YZY second-harmonic generation (SHG) in PPKTP (the first letter denotes the polarization direction of the second harmonic field with respect to the crystal axes and the last two letters denote the polarization of the fundamental fields). It has been shown that, when used in an OPO, such a nonlinear medium would generate tripartite continuous-variable entanglement below\cite{us} and also {\em above}\cite{ashton} threshold. Figure \ref{qpm} illustrates the theoretical prediction of QPM-SHG for propagation along the x-axis of PPKTP using the most recent Sellmeier coefficients.\cite{kato,emanueli} 
%-------------
%-------------
For a polarization set $ijk$, the required poling period $\Lambda_{ijk} = \lambda/[2n_i(\lambda/2)-n_j(\lambda)-n_k(\lambda)]$ is plotted versus the fundamental wavelength $\lambda$, at 40$^{\circ}$ C, and the crossing points, which indicate coincidences, are marked by circles for ref. \cite{kato} and by squares for ref. \cite{emanueli}. Because the poling is a square spatial modulation, one has also access to harmonics of the Fourier poling vector, i.e.\ subharmonics of the poling period. (We realized in the course of this study that related work, at shorter wavelengths and with no quantum information connections, was done independently by Pasiskevicius {\em et al.}\cite{laurell}.) As can be seen from Fig.~\ref{qpm}, the Sellmeier coefficients of Ref.~\cite{emanueli} predict $\Lambda_{YZY}=2\Lambda_{ZZZ} \simeq 48 \mu$m at $\lambda \simeq 1.53 \mu$m and $2\Lambda_{ZZZ}=7\Lambda_{ZYY}\simeq 45 \mu$m at $\lambda\simeq 1.485 \mu$m, whereas those of Ref.~\cite{kato} predict a rather remarkable triple coincidence  $\Lambda_{YZY}=2\Lambda_{ZZZ}=7\Lambda_{ZYY}\simeq 45.5 \mu$m at $\lambda\simeq 1.49 \mu$m,  all in the low-loss window of silica optical fibers.

\begin{figure}[htb]
\begin{center}
\includegraphics[width=3in]{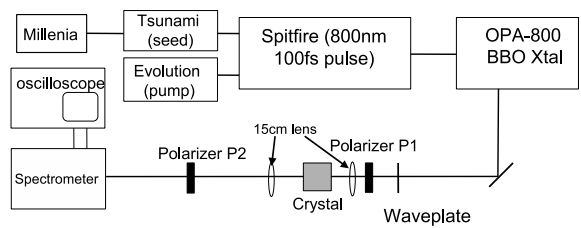}
\end{center}
\caption{Experimental setup.  The polarizer P1 and waveplate combination were used to adjust input power and polarization, while the polarizer P2 was used to analyze the output incident on the spectrometer.  A Si photodiode detected the ouput of the spectrometer which was set to scan over the wavelength range of interest.}
\label{setup}
\end{figure}

\begin{figure}[htb]
\begin{center}
\includegraphics[width=3in]{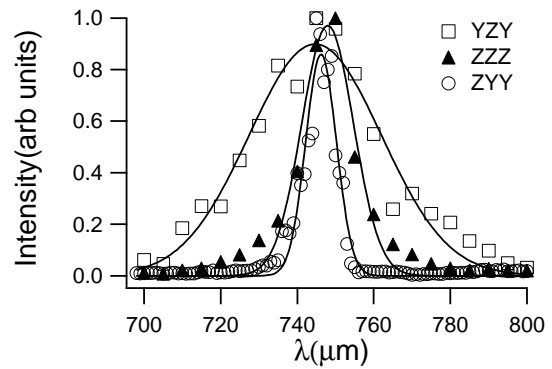}
\end{center}
\caption{Output SHG powers of the three coincident nonlinear processes at 22 $^{\circ}$ C, normalized to the same scale (slightly different total input powers were used in each case). Gaussian curves were fitted to the data to determine the center wavelength of each signal (see text).}
\label{crossing}
\end{figure}

We therefore had to choose which prediction to test and decided to start with the triple coincidence. We used an X-cut, 7 mm long, 2 mm $\times$ 1 mm section PPKTP crystal, poled with a 45.65 $\mu$m period by Raicol Crystals, Ltd. The crystal was pumped by 150 fs pulses at 1490 nm, which were focused to a waist of 55 $\mu$m  by a 15 cm focal length lens. The input light was generated by a Spectra-Physics OPA-800. A BBO crystal inside the OPA was angle tuned in order to select the peak signal wavelength.  All other wavelengths were blocked by a polarizer before the PPKTP crystal.  The OPA-800 was pumped by a Spectra-Physics Spitfire regenerative amplifier with 800 nm wavelength and 100 fs pulse length at a 1 kHz repetition rate (Fig.~\ref{setup}).
%-------------
%-------------
The PPKTP crystal was placed in an oven, temperature controlled to less than $\rm 0.1^o$ C. The output beams of the crystal were sent to a computer-controlled Acton Research SprectraPro 500i grating monochromator. The combination or the P1 and P2 polarizers allowed selection of any desired phase matching conditions in the crystal.  To obtain a signal for YZY phase matching, the waveplate and P1 were arranged to set the input polarization at $\rm 45^o$ from Y and Z, and P2 was rotated to verify there was no signal when blocking Y-polarized light and passing Z-polarized light, thus verifying that no additional unexpected phase-matching conditions had been met inside the crystal. The expected signal was then observed by rotating P2 to allow the Y-polarized light through to the spectrometer.  This method was used to analyze each of the signal spectra, so that it could be definitively determined that each signal corresponded to a very specific phase matching condition, defined by the pump and signal polarizations. The experimental results 
are displayed in Fig.~\ref{crossing}.
%-------------
%-------------
The pump energy per pulse was typically 10 $\mu$J, corresponding to a peak power of 67 MW per pulse. Thus, the fundamental pump beam was significantly depleted (30\%-50\%). The different widths of the SHG peaks are related to all possible SHG and sum-frequency generation (SFG) processes within the pump's spectral width (50 nm FWHM), the modeling of which is involved in the depleted pump regime. However, the crucial point here is that the 3 peak centers are all located within a few nanometers: at 747.8 nm for ZZZ, 745.4 nm for YZY, and 745.8 nm for ZYY, to be compared with the predicted values (Fig.~\ref{qpm}) 744.3 nm for ZZZ, 746 nm for YZY, and 742.8 nm for ZYY. (A systematic error of 1.2nm is associated with each peak center).  This therefore confirms the analysis of Ref. \cite{kato}, since Ref. \cite{emanueli} predicts the ZZZ-YZY coincidence should not be observed. We also observed third-harmonic generation of blue light from cascading of SHG with SFG from the fundamental frequency. We have not investigated the efficiency of the SFG process, however, as it is rather weak and is predicted to overlap only partially with SHG. 

\begin{figure}[htb]
\begin{center}
\includegraphics[scale=.4]{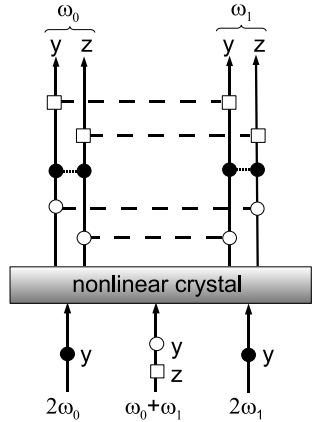}
\end{center}
\caption{A quadripartite CV entangler based on the triple concurrence observed in this work.  The horizontal separation of input and output beams refers to their frequencies, not their wave vectors, which all overlap. A frequency scale factor of 2 has been applied between the input (pump) and output (signal) OPO beams. As previously, letters $y$ and $z$ denote the optical polarization direction of the concerned field with respect to the crystal's axes.}
\label{entscheme}
\end{figure}

These results provide the possibility of designing a quadripartite entanglement source even simpler and more compact than the one we proposed in Ref.~\cite{us}. The principle is depicted in Fig.~\ref{entscheme}. An amplitude-modulated beam can provide the 3 pump fields at frequencies $2\omega_0$, $\omega_0+\omega_1$, and $2\omega_1$ that will yield concurrent photon pairs into the four entangled OPO modes.
%-------------
%-------------
Note that each pair emission process constitutes a well known CV bipartite entanglement process in itself\cite{reid,ou} and that the overall multipartite entanglement results from the {\em coherent concurrence} of all these bipartite entanglers.\cite{us}

In conclusion, we have observed a triple coincidence between different harmonics of all allowed QPM processes when propagating along the X axis in PPKTP. This result confirms the theoretical analysis of KTP Sellmeier coefficients by Kato and Takaoka\cite{kato} and paves the way to the realization of compact sources of tripartite and quadripartite CV GHZ states. Retaining the triple coincidence in the narrowband continuous-wave (CW) regime may be difficult since the crossing is not exact any more at high spectral resolution (Fig.~\ref{qpm}) and temperature tuning is only enough of a constraint to guarantee a double coincidence. However, we can still have the lines in Fig.~\ref{qpm} cross perfectly by using longer pulses of a few tens of ps duration (a few nm resolution). Then the scheme of Fig.~\ref{entscheme} is quite feasible. We are now proceeding to use PPKTP in an OPO cavity to demonstrate the compact generation of bright tripartite GHZ states. Note, again, that the entanglement wavelength is ideally suited for quantum communication using optical fibers.

Part of this work was done at the ultrafast laser laboratory of the Center for Atomic, Molecular, and Optical Science at the University of Virginia. We thank its Director, Kurt Kolasinski, for his support and technical help. We also thank Kristy DeWitt for help with the pulsed OPA source. Finally we thank Alex Skliar, from Raicol Crystal Ltd., for his help with PPKTP. This work is supported by NSF grant PHY-0240532 and by the NSF IGERT SELIM program at the University of Virginia.

\end{document}